\documentclass[reqno]{amsart}

\usepackage{graphicx}

\theoremstyle{plain}
\newtheorem{theorem}{Theorem}

\newtheorem{proposition}{Proposition}
\newtheorem{lemma}{Lemma}
\newtheorem{corollary}{Corollary}
\theoremstyle{definition}

\usepackage{amsfonts,amsmath}

\def\R{{\mathbb R}}
\def\C{{\mathbb C}}
\def\mod{{\mathrm{mod}}\,}

\begin{document}
%
%
\title[Moutard transformation: an algebraic formalism and applications]{The Moutard transformation: an algebraic formalism
via pseudodifferential operators and applications}
\author[I.A. Taimanov]{Iskander A. Taimanov}
\address[I.A. Taimanov]{Institute of Mathematics,
630090 Novosibirsk, Russia} \email {taimanov@math.nsc.ru}

\author[S.P. Tsarev]{Sergey P. Tsarev}
\address[S.P. Tsarev]{Siberian
Federal University, Svobodnyi avenue, 79, 660041, Krasnoyarsk,
Russia} \email{sptsarev@mail.ru}
\date{\today}
%
\begin{abstract}
In this paper we consider the Moutard transformation \cite{mou}
which is a two-dimensional version of the well-known Darboux
transformation. We give an algebraic interpretation of the Moutard
transformation as a conjugation in an appropriate ring and the
corresponding version of the algebro-geometric formalism for
two-dimensional Schr\"odinger operators. An application to some
problems of the spectral theory of two-dimensional Schr\"odinger
operators and to the $(2+1)$-dimensional Novikov--Veselov equation
is sketched.
\end{abstract}
%
\thanks {This research was partially supported by
Siberian Branch of RAS (the interdisciplinary integration project
No. 65) (the first author) and the RFBR Grants 09-01-00762-a and
06-01-89507-NNS-a (the second author).}

\maketitle


\section*{Introduction}
\label{sec-dm}

The Moutard transformation plays a fundamental role in
projective--differential geometry of surfaces  and was extensively
investigated by  Bianchi, Darboux, Demoulin, Guichard and others.

In recent publications \cite{TT07,TT08,TT-DAN08} we gave an
application of the Moutard transformation to the explicit
construction of two-dimensional Schr\"odinger operators
$$
H = -\Delta + u= -(\partial_x^2 +\partial_y^2) + u(x,y)
$$
with fast decaying smooth rational potentials such that their
$L_2$-ker\-nels  contain at least two-dimensional subspaces spanned
by rational eigenfunctions as well as a $(2+1)$-dimensional
extension of the Moutard transformation which was able to produce
explicit rational blowing-up solutions to the Novikov--Veselov (NV)
equation with fast decaying smooth rational Cauchy data.

Purely algebraic constructions of the aforementioned papers imply
that one should look for an algebraic interpretation of the Moutard
transformation as a simple transformation in the ring of partial
differential operators. The one-dimensional case of Darboux
transformation was studied in \cite{EK}. Here we give an analogue of
their results; as it turned out, for the Moutard transformation one
needs to recourse to a more complicated ring of \emph{rational}
pseudodifferential operators (Ore localization of the ring of
partial differential operators). As a by-product, we obtain a more
natural version of the Dubrovin-Krichever-Novikov formalism
\cite{DKN,K76} of algebro-geometric two-dimensional Schr\"odinger
operators. In the last Section we briefly sketch our previous
results \cite{TT07,TT08,TT-DAN08}.

First we give an account of the algebraic theory of the Darboux
transformations.

\subsection{The Darboux transformation} Let
$$
H = -\frac{d^2}{dx^2} + u(x)
$$
be a one-dimensional Schr\"odinger operator and let $\omega$ satisfy
the equation
$$
H \omega = 0.
$$
The function $\omega$ determines a factorization of $H$:
\begin{equation} \label{onefactor} H = A^\top A, \ \ \ A =
-\frac{d}{dx} + v, \ \ \ A^\top = \frac{d}{dx}+v, \ \ \ v =
\frac{\omega_x}{\omega}.
\end{equation}
 Indeed we have
$$
A^\top A = \left(\frac{d}{dx} + v\right)\left(-\frac{d}{dx} +
v\right) = -\frac{d^2}{dx^2} + v^2 + v_x
$$
and the equation
$$
v_x + v^2 = u
$$
is equivalent to $H\omega = 0$. If $v$ is real-valued we have
$A^\ast = A^\top. $

{\it The Darboux transformation}\footnote{Sometimes it is also
called the Crum transformation due to \cite{Crum}. Burchnall and
Chaundy called it transference \cite{BC}.} is the swapping of
$A^\top$ and $A$:
$$
H = A^\top A \to \widetilde{H} = AA^\top,
$$
or in terms of $u$:
$$
u = v^2 + v_x  \to \widetilde{u} = v^2 - v_x.
$$

It is easy to check by simple computations that

\begin{proposition}
\label{prop1} If $\varphi$ satisfies the equation $H \varphi = E
\varphi$ with $E = \mathrm{const}$ then $\widetilde{\varphi} = A
\varphi$ satisfies the equation $\widetilde{H} \widetilde{\varphi} =
E \widetilde{\varphi}.$
\end{proposition}

{\sc Remark.} In general the Darboux transformation is defined for
any solution to the equation $H \omega = c\omega$ with
$c=\mathrm{const}$ (see, for instance, \cite{EK}). In this case it
reduces to the transformation of $H^\prime = H-c$ for which
$H^\prime\omega = 0$.

\subsection{The Moutard transformation}

This transformation was invented by Th.~Moutard for the hyperbolic
equation of the form $\psi_{xy}-u(x,y)\psi=0$ in the context of
local differential geometry of surfaces and was extensively studied
and used in many areas of the local surface theory \cite{Dar}. Here
we give an elliptic version \cite{TT07,TT08,TT-DAN08} of this
transformation suitable for our purposes.

Let $H$ be a two-dimensional potential Schr\"odinger operator and
let $\omega$ be a solution to the equation
$$
H \omega = (- \Delta + u )\omega = 0.
$$
Then {\sc the Moutard transformation} of $H$ is defined as
$$
\widetilde{H} = -\Delta + u -2\Delta \log \omega = -\Delta - u +
2\frac{\omega_x^2+\omega_y^2}{\omega^2}.
$$

\begin{proposition}
\label{prop3} If $\varphi$ satisfies the equation $H \varphi = 0$,
then the function $\theta$ defined from the system
\begin{equation}
\label{eigen-moutard} (\omega \theta)_x = -\omega^2
\left(\frac{\varphi}{\omega}\right)_y, \ \ \ (\omega \theta)_y =
\omega^2 \left(\frac{\varphi}{\omega}\right)_x
\end{equation}
satisfies $\widetilde{H}\theta = 0$.
\end{proposition}

By this definition, if $\theta$ satisfies (\ref{eigen-moutard}) then
\begin{equation} \label{theta-family}
 \theta + \frac{t}{\omega}, \ \ \ t =
\mathrm{const},
\end{equation}
satisfies (\ref{eigen-moutard}) for
any constant $t$.

We shall use the following notation for the Moutard transformation:
$$
M_\omega(u) = \widetilde{u} = u - 2\Delta \log \omega, \ \ \
M_\omega(\varphi) = \{\theta+\frac{t}{\omega}, \ t \in \C\}.
$$

For one-dimensional potentials the Moutard transformation reduces to
the Darboux transformation. Indeed, let $u=u(x)$ depend on $x$ only
and $\omega=f(x)e^{\sqrt{c}y}$. Then $f$ satisfies the
one-dimensional Schr\"odinger equation
$$
H_0 f = \left( - \frac{d^2}{dx^2} + u\right) f = cf
$$
and the Moutard transformation reduces to the Darboux transformation
of $H_0$ defined by $f$:
$$
H = H_0  - \frac{\partial^2}{\partial y^2} \ \ \longrightarrow \ \
\widetilde{H} = \widetilde{H_0} -\frac{\partial^2}{\partial y^2}.
$$
If $g=g(x)$ satisfies $H_0 g = E g$, then $H \varphi = 0$ with
$\varphi = e^{\sqrt{E}y}g(x)$. We derive from (\ref{eigen-moutard})
that $\theta = e^{\sqrt{E}y}h(x)$ satisfies $\widetilde{H}\theta =
0$ if $h = \frac{1}{\sqrt{c}+\sqrt{E}} \left(\frac{d}{dx} -
\frac{f_x}{f}\right)g$, i.e. $h$ is a multiple of the Darboux
transform of $g$: $h =  - \frac{1}{\sqrt{c}+\sqrt{E}}Ag$ where $H_0
- c = A^\top A$ is the factorization of $H_0-c$ defined by $f$. The
inverse Darboux transformation is given by $g =
\frac{1}{\sqrt{c}-\sqrt{E}}\left( \frac{d}{dx} +
\frac{f_x}{f}\right)h$.

{\sc Remark.} We can rewrite (\ref{eigen-moutard}) as
\begin{equation} \label{eigen-moutard2} \left(\bar{\partial} +
\frac{\omega_{\bar{z}}}{\omega}\right) \theta = i
\left(\bar{\partial} - \frac{\omega_{\bar{z}}}{\omega}\right)
\varphi, \ \ \ \left(\partial + \frac{\omega_z}{\omega}\right)
\theta = -i \left(\partial - \frac{\omega_z}{\omega}\right) \varphi
\end{equation}
which implies
\begin{equation} \label{dress-moutard}
\omega^{-1}\cdot \bar{\partial}\cdot \omega (\theta) = i \omega
\cdot \bar{\partial}\cdot \omega^{-1}(\varphi), \ \ \omega^{-1}\cdot
\partial\cdot \omega (\theta) = - i \omega \cdot \partial \cdot
\omega^{-1}(\varphi),
\end{equation}
where, as usual, $z=x+i\,y$,
$\partial=\partial_z=\frac{1}{2}(\partial_x- i
\partial_y)$, $\bar\partial=\partial_{\bar z}=\frac{1}{2}(\partial_x + i \partial_y)$.
This representation will be used later.

There is another two-dimensional generalization of the Darboux
transformation called the Laplace transformation. Its relation to
integrable systems was recently studied in \cite{NV}.

\section{The spectral curves of algebraic Schr\"odinger operators}

Since weakly algebraic two-dimensional Schr\"odinger operators are
defined in terms of their spectral curves on the zero energy level
we recall the definitions of the spectral curves of one- and
two-dimensional Schr\"odinger operators.

\subsection{One-dimensional Schr\"odinger operators} Let a
one-dimensional Schr\"odinger operator
$$
H = -\frac{d^2}{d x^2} +u(x)
$$
commute with an ordinary differential operator $L$ of order $2n+1$.
Then the Burchnall--Chaundy theorem \cite{BC} guarantees that these
two operators satisfy a polynomial equation
$$
Q(H,L) = 0, \ \ \ Q(\lambda,E) = \lambda^2 - P_{2n+1}(E) = 0,
$$
where $P_{2n+1}(E)$ is a polynomial of degree $2n+1$. This equation
defines {\it the spectral curve} $\Gamma$ of the operator $H$. Its
points parameterize the joint ``eigenfunctions'', i.e. solutions of
the equations
$$
H\psi = E\psi,  \ \ \ L\psi = \lambda \psi.
$$
Moreover these functions are glued (after some normalization) into a
function $\psi(P,x)$ which is meromorphic in $P \in \Gamma$ and
after completing $\Gamma$ by adding an infinity point $E=\infty$ the
eigenfunction $\psi$ gets a singularity at $\infty$:
$$
\psi(P,x) \approx e^{ikx}
$$
where $k^{-1} = \frac{1}{\sqrt{E}}$ is a local parameter on $\Gamma$
near the infinity point $\infty$ \cite{N,DMN}. Therewith it is said
that the operator $H$ is {\it algebraic} (or {\it
algebro-geometric}).

{\sc Example.} {\it The spectral curve of a constant potential.} Let
$u(x)= c$ be a constant potential, then $\Gamma$ is given by
$\lambda^2 = E-c$,
$$
\psi_{\pm} = e^{\pm i\lambda x},
$$
and the covering $(\lambda,E) \to E$ ramifies at $E=c$ where
$\psi_\pm(x,c)=1$.

\subsection{Two-dimensional Schr\"odinger operators} Given a
partial differential operator $H$, one says that operators $L_1,
\dots, L_n$ generate a commutative $(\mod H)$ algebra ${\mathcal A}$
if they satisfy the following commutation relations:
$$
[L_i,L_j]=D_{ij}H, \ \ \ [L_i,H] = D_i H
$$
where $D_{ij}, D_i, 1 \leq i,j \leq n$, are partial differential
operators. This definition was introduced in \cite{DKN} in which
they considered in detail the particular case when $H$ is the
two-dimensional Schr\"odinger operator (probably with an
electromagnetic field) in two variables and introduced the following
definition:

\begin{itemize}
\item
a two-dimensional operator
$$
H = \partial\bar{\partial}+ v \bar{\partial} + u
$$
is called {\it (weakly) algebraic} if it is included in a nontrivial
commutative $(\mod H)$ algebra ${\mathcal A}$ generated by operators
$L_1$ and $L_2$ in two variables, the operators $L_1$ and $L_2$
satisfy the polynomial relation
$$
Q(L_1,L_2) = 0 (\mod H)
$$
and to a generic point of the algebraic curve
$Q(\lambda_1,\lambda_2)=0$ there corresponds a $k$-dimensional space
of functions $\psi$ which satisfy the equations
$$
H\psi =0, \ \ \ L_i \psi = \lambda_i \psi, \ i=1,2.
$$
The dimension $k$ is called the rank of ${\mathcal A}$.
\end{itemize}

\section{The conjugation representation}

\subsection{The Darboux transformation}

The Darboux transformation admits the well-known representation as a
conjugation in the ring of pseudodifferential operators which
implies interesting corollaries concerning algebraic operators.

Let $H \omega = 0$ and $A = -\frac{d}{dx} =
\frac{\omega_x}{\omega}$. Then in the ring of pseudodifferential
operators in $x$ we have
$$
A = - \omega \cdot \frac{d}{dx} \cdot \omega^{-1}, \ \ A^\top =
\omega^{-1} \cdot \frac{d}{dx} \cdot \omega, \ \ H = A^\top A,
$$
where the functions $\omega$ and $\omega^{-1}$ are identified with
the operator of multiplication by them. From that we conclude

\begin{proposition}
\begin{equation} \label{darboux1}
\widetilde{H} = A \cdot H \cdot A^{-1}.
\end{equation}
\end{proposition}

Fixing $\omega$ let us denote by
$$
\widetilde{M} = A \cdot M \cdot A^{-1}
$$
the conjugation of a pseudodifferential operator $M$. It is clear
that if $L$ commutes with $H$, then $\widetilde{L}$ commutes with
$\widetilde{H}$:
$$
[H,L]= 0 \ \ \ \Rightarrow \ \ \ [\widetilde{H},\widetilde{L}]=0.
$$
However, given an arbitrary differential operator $M$, we have
$$
M \cdot \omega = M\omega + M^\prime \cdot \frac{d}{dx}
$$
where $M\omega$ is the function obtained by applying $M$ to $\omega$
and $M^\prime$ is a differential operator. Hence
$$
\widetilde{M} = \omega \cdot \frac{d}{dx} \cdot \omega^{-1} \cdot M
\cdot \omega \cdot \left(\frac{d}{dx}\right)^{-1} \cdot \omega^{-1}
=
$$
$$
\omega \cdot \frac{d}{dx} \cdot \omega^{-1} \cdot M\omega \cdot
\left(\frac{d}{dx}\right)^{-1} \cdot \omega^{-1} + \omega \cdot
\frac{d}{dx} \cdot \omega^{-1} \cdot M^\prime \cdot \omega^{-1}
$$
and we conclude that

{\sl $\widetilde{M}$ is a differential operator if and only if
$\omega^{-1} \cdot M\omega = \lambda = \mathrm{const}$, which means
$$
M\omega = \lambda \omega.
$$
}

Let $H\omega=0$ and let $L$ be a differential operator of odd order
which commutes with $H$. We assume that its order is minimal with
respect to this property. Let, by the Burchnall--Chaundy theorem,
$$
Q(H,L) = 0, \ \ \ Q(E,\lambda) = \lambda^2 - P(E).
$$
Then we have (see \cite{EK}):

\begin{enumerate}
\item
if $L \omega =\lambda\omega$ and $P(E)$ has no a multiple root at
$E=0$, then $\widetilde{H}$ and $\widetilde{L}$ generate a
commutative ring of differential operators and
$$
Q(\widetilde{H},\widetilde{L}) = 0;
$$

\item
if $\omega^{-1} \cdot L\omega \neq \mathrm{const}$, then
$\widetilde{H}$ and $L_0 = \widetilde{LH}$ generate a commutative
ring of differential operators and
$$
Q_+(\widetilde{H},L_0) = 0, \ \ \ Q_+(E,\lambda) = \lambda^2 -
E^2P(E);
$$

\item
if $L \omega =\lambda\omega$ and $P(E)$ has a multiple root at
$E=0$, then the action of the Darboux transformation is inverse to
one described in the previous statement: $\widetilde{H}$ and some
operator $L_0$ generate commutative ring of differential operators
with
$$
Q_-(\widetilde{H},L_0) = 0, \ \ \ Q_-(E,\lambda) = \lambda^2 -
E^{-2}P(E).
$$
\end{enumerate}

In particular, this implies that

\begin{itemize}
\item
The Darboux transformation preserves the class of algebro-geometric
one-di\-men\-si\-o\-nal Schr\"odinger operators. Moreover it always
preserves the normalization of the spectral curve.
\end{itemize}

In \cite{EK} it is noted that some facts mentioned above were
discovered by Burchnall, Chaundy and Drach.

The first two cases are demonstrated by the following

{\sc Example.} {\it The Darboux transformation of the constant
potential.} Let $u =c = \mathrm{const}$ and $L = \frac{d}{dx}$.

\begin{enumerate}
\item
Let $\omega = e^{\sqrt{c-c_0}x}$. Then $v = \frac{\omega_x}{\omega}
= \sqrt{c-c_0}$, $v_x=0$, and the Darboux transformation is even
trivial.

\item
Let $\omega = \frac{1}{2}(e^{\sqrt{c-c_0}x}+e^{-\sqrt{c-c_0}x}) =
\cos (\sqrt{c-c_0}x)$. Then
$$
v = - \sqrt{c-c_0} \tan (\sqrt{c-c_0}x), \ \ \ v_x =
\frac{c_0-c}{\cos^2 (\sqrt{c-c_0}x)},
$$
$$
u=c \to \widetilde{u} = c + \frac{2(c-c_0)}{\cos^2 (\sqrt{c-c_0}x)}.
$$
If $c-c_0 <0$, then $\widetilde{u}$ is not periodic.
\end{enumerate}

An example of the third case is  easily derived from the following
explicitly computable examples which are interesting in themselves.

{\sc Example.} {\it Rational solitons via the Darboux
transformation.} Let $u =0$ and $\omega = x$. Then
$$
v = \frac{1}{x}, \ \ \ v_x = - \frac{1}{x^2}, \ \ \ \widetilde{u} =
\frac{2}{x^2}.
$$
The spectral curve $\Gamma = \{\lambda^2 = E\}$ is transformed to
$\widetilde{\Gamma} = \{\lambda^2 = E^3\}$ and
$$
\psi = \left(1 - \frac{1}{i\sqrt{E}x}\right)e^{i\sqrt{E}x}
$$
(we normalize it by condition $\psi \approx e^{i\sqrt{E}x}$ as $E
\to \infty$). The iterations of the Darboux transformation initially
applied to the trivial potential $u=0$ give all rational solitons discovered in \cite{AMM}. The
spectral curve of the potential
$$
u_n = \frac{n(n+1)}{2x^2},
$$
obtained after $n$ iterations, is given by the equation $\lambda^2 =
E^{2n+1}$. The spectral curves of these are singular: topologically
they are spheres but at $E=0$ they have the following singularities:

{\it any rational function $f$ on $\Gamma_n$, the spectral curve of
$u_n$, which is holomorphic near $E=0$ satisfies the condition}
\begin{equation}
\label{singular-1} f^\prime = f^{\prime\prime\prime} = \dots =
f^{(2n-1)}=0.
\end{equation}

\noindent This is easily explained by the normalization mapping $\C
\to \Gamma= \{\lambda^2=E^{2n+1}\}$ which has the form
$$
t \to (\lambda = t^{2n+1}, E = t^2).
$$
Indeed any rational function $f$ which is holomorphic at $E=0$ is a
ratio of polynomials $\frac{P(\lambda,E)}{Q(\lambda,E)}$ such that
$Q(0,0) \neq 0$ and any such function written in terms of $t$
satisfies the conditions (\ref{singular-1}).

\subsection{The Ore localization}
\label{sec-ore}

In order to give an analogous interpretation of the Moutard
transformation as a conjugation we first recall some basic facts
from the Ore theory of localization of noncommutative rings and
introduce the rings $F(\partial_y)$ and $F(\partial_y)[\partial_x]$
which we shall use.

We recall that a ring $R$ is an algebra with two operations, the
addition and the multiplication, such that $R$ is a commutative
group with respect to the addition and the distribute laws
$$
a(b+c) = ab+ ac, \ \ \ (a+b)c = ac + bc \ \ \ \mbox{for all $a,b,c
\in R$}
$$
hold. It is said that a ring $R$ is regular if it satisfies {\it the
Ore conditions}:

\begin{itemize}
\item
for $a \neq 0, b \neq 0$ there exist $r \neq 0, s \neq 0$ such that
$ar = bs$;

\item
if $ab=ac$ or $ba=ca$ for $a \neq 0$ then $b=c$ (so it has no zero
divisors).
\end{itemize}

Ore showed in \cite{Ore} that

{\sl any regular ring can be embedded as a subring into a
non-commutative field (skew field of fractions)}

\noindent as follows. Let us consider the set $S$ of all formal
fractions
$$
ab^{-1} = \left(\frac{a}{b}\right), \ \ \ b \neq 0.
$$
\emph{Note} that due to non-commutativity of $R$ one shall always
keep in mind that in our construction the denominators $b^{-1}$ are
always on the \emph{right} of the numerators $a$! It is easy to
propose an analogous construction with denominators standing on the
left, but the resulting skew field will be isomorphic to the field
we are to construct.

We say that two such fractions are equal:
$$
\left(\frac{a}{b}\right) = \left(\frac{c}{d}\right)
$$
if and only if
$$
ar = cs
$$
where $r$ and $s$ satisfy the equality (see the first Ore condition)
\begin{equation} \label{ore-pair}
br = ds.
\end{equation}
It is easy to show that the equality is independent on the choice of
$r$ and $s$ satisfying the last equality. Then the addition is
defined by
$$
\left(\frac{a}{b}\right) + \left(\frac{c}{d}\right) = \left(
\frac{ar+cs}{br}\right) = \left(\frac{ar+cs}{ds}\right)
$$
where $r$ and $s$ satisfy (\ref{ore-pair}). The multiplication is
given by
$$
\left(\frac{a}{b}\right)\left(\frac{c}{d}\right) =
\left(\frac{at}{du}\right)
$$
where
$$
bt = au.
$$
The unit element is defined as
$$
\left(\frac{a}{a}\right) = 1, \ \ \ a \neq 0,
$$
this definition is independent on the choice of $a$ and with such a
law of multiplication we have
$\left(\frac{a}{b}\right)\left(\frac{b}{a}\right) = 1$.

We say that $S$ (with these operations) is {\it the Ore
localization} of $R$. One can routinely check all the usual
properties of the addition and multiplication operations defined on
$S$, which becomes a \emph{skew field}. This means that we can use
expressions like $A^{-1}$ for arbitrary elements of this skew field.
All operations in $S$ given above are constructive, unlike the case
of the ring of formal pseudodifferential operators defined as
infinite series customary in solitonics.

For a commutative ring $R$ without divisors of zero the Ore
localization coincides with the standard localization. Let us
introduce the main example with non-commutative ring $R$ which we
shall use.

{\sc Example.} Let $F = {\bf k}(x_1,\dots,x_n)$ be a field formed by
all ${\bf k}$-valued functions of the form
$$
\frac{P(x_1,\dots,x_n)}{Q(x_1,\dots,x_n)}
$$
where $P$ and $Q$ are analytical functions of $x_1,\dots,x_n$ or
even formal power series in these variables. We consider two cases
which are

1) ${\bf k} = \R$ and $x_1,\dots,x_n \in \R$;

2) ${\bf k} = \C$, $n=2m$ and $(x_1,\dots,x_n) =
(z_1,\bar{z}_1,\dots,z_m,\bar{z}_m)$,

\noindent and, for simplicity and our needs, we even assume that
$n=2$, $x_1 = x$, and $x_2 = y$. Let us consider $F$ as an operator
algebra $R$ which acts on itself by multiplications
$$
f(g) = fg, \ \ \ f , g \in F,
$$
take the partial derivative operators $\partial_x =
\frac{\partial}{\partial x},
\partial_y = \frac{\partial}{\partial y}$
and consider the operator algebra $F[\partial_y]$ which is generated
by elements of $R$ and by the operator $\partial_y$ that acts on
$F$. The ring  $F[\partial_y]$ is noncommutative because
\begin{equation}
\label{comm} [\partial_y,f] = \partial_y \cdot f - f \cdot
\partial_y = \frac{\partial f}{\partial y} =  f_y \ \  \mbox{for $f
\in F$}.
\end{equation}
It is straightforward to check

\begin{proposition}
$F[\partial_y]$ and $F[\partial_x,\partial_y]$ satisfy the Ore
conditions.
\end{proposition}

Let us take the Ore localization $F(\partial_y)$ of $F[\partial_y]$.
We have

\begin{proposition}
\label{commutator} For any $f \in F$ we have
$$
[\partial_y^{-1},f] = - \partial_y^{-1}\cdot \frac{\partial
f}{\partial y}\cdot
\partial^{-1}_y.
$$
\end{proposition}

{\sc Proof.} Let us take the equality (\ref{comm}), multiply every
its term by $\partial_y^{-1}$ from both sides and obtain
$[f,\partial_y] = \partial_y^{-1} \cdot f_y \cdot \partial_y^{-1}$.
Proposition is proved.

Let us consider $F(\partial_y)[\partial_x]$, the ring of formal
differential operators in $x$ with coefficients from
$F(\partial_y)$. This ring is embedded into its Ore localization
$F(\partial_x,\partial_y)$.

We need to define the commutation of $\partial_x^{\pm 1}$ with
elements of $F(\partial_y)[\partial_x]$. Since
$$
[\partial_x^{-1},f] = -
\partial_x^{-1}[\partial_x,f]\partial_x^{-1},
$$
it is enough to define the commutators of $f \in
F(\partial_y)[\partial_x]$ with $\partial_x$. Here we do that even
for the Ore localization ring $F(\partial_x,\partial_y)$.

If $L \in F[\partial_x,\partial_y]$ is a differential operator in
$x$ and $y$ with coefficients from $F$, then its derivative in $x$
(i.e.\ the operator with coefficients being the derivatives of the
respective coefficients of $L$) may be defined by the commutation
formula
$$
L_x = [\partial_x, L].
$$
Let us take this formula as a definition of the derivative of any
element of $P \in F(\partial_x,\partial_y)$ w.r.t.\ $x$. Using the
Leibnitz identity
$$
(LM)_x = L_xM + LM_x, \ \ \ L,M \in F(\partial_x,\partial_y),
$$
we derive the following proposition by straightforward computations.

\begin{proposition}
Given $P = M \cdot L^{-1}$ with $L, M \in F[\partial_x,\partial_y]$,
we have
$$
P_x \stackrel{\mathrm{def}}{=} [\partial_x,P] = M(L^{-1})_x + M_x
L^{-1} = -M\,L^{-1}L_xL^{-1} + M_x L^{-1}.
$$
\end{proposition}

\begin{corollary}
\label{cor1} If $P \in F(\partial_y)$, then $P_x = [\partial_x,P]
\in F(\partial_y)$, i.e. $P_x$ contains no derivations in $x$.
\end{corollary}

\subsection{The Moutard transformation}

Let us consider the general Moutard transformation which is applied
to an operator:
$$
H=\partial_r\partial_s - u(r,s)
$$
where
$$
r=x, \ \ s = y, \ \ x,y \in \R \ \ \ \mbox{(the hyperbolic case)},
$$
or
$$
r = z, \ \ s = \bar{z}, \ \ z \in \C \ \ \ \mbox{(the elliptic
case).}
$$
Via the formulas
\begin{equation}
\label{M0} (\omega \theta)_r = - \omega^2
\left(\frac{\varphi}{\omega}\right)_r, \ \ \ (\omega\theta)_s =
\omega^2\left(\frac{\varphi}{\omega}\right)_s,
\end{equation}
it relates solutions $\varphi$ and $\theta$ to the equations
$$
H\varphi =0, \ \ \ \widetilde{H} \theta =0
$$
where
$$
\widetilde{H} = \partial_r\partial_s - \widetilde{u}, \ \
\widetilde{u} = u - 2\partial_r\partial_s \log \omega= - u + 2
\frac{\omega_r\omega_s}{\omega^2}.
$$

Let us consider the Ore localization $F(\partial_s)$ of the ring of
differential operators in $s$ and the ring
$F(\partial_s)[\partial_r]$ of differential operators in $r$ with
coefficients from (noncommutative) ring $F(\partial_s)$. Given
$\omega$, a solution to the equation $H\omega=0$, we consider the
differential operators $A,B \in F[\partial_s]$:\
$$
A=\omega^{-1}\cdot\partial_s\cdot\omega, \ \ \ B=\omega\cdot
\partial_s\cdot{\omega^{-1}}
$$
and their ``ratio'', the operator $\Omega$ of the form
$$
\Omega = A^{-1}\cdot B =
\Big(\omega^{-1}\cdot\partial_s\cdot\omega\Big)^{-1}\cdot\Big(\omega\cdot
\partial_s\cdot{\omega^{-1}}\Big)=A^{-1} B.
$$
We denote by $M$ and $\widetilde{M}$ the following formal operators
of the first order from $F(\partial_s)[\partial_r]$:
$$
M = \partial_s^{-1}\cdot H =
\partial_r -  \partial_s^{-1}\cdot u,
\qquad \widetilde M = \partial_s^{-1}\cdot \widetilde{H} =
\partial_r - \partial_s^{-1}\cdot \widetilde{u}.
$$

\begin{theorem}\label{th1}
Given $\omega$ satisfying $H\omega = (\partial_r\partial_s +
u)\omega=0$ and the Moutard transform $\widetilde{u}$ of $u$ (we
assume that the transformation is generated by $\omega$), the
operators $M$ and $\widetilde{M}$ are conjugated in
$F(\partial_s)[\partial_r]$ by $\Omega$:
$$
\widetilde{M} = \Omega \cdot M \cdot \Omega^{-1}.
$$
\end{theorem}

{\sc Remark.}  It is easy to check that $H$ and $\widetilde{H}$ are
not conjugated by $\Omega$: $ \widetilde{H} \neq \Omega \cdot H
\cdot \Omega^{-1}$.

{\sc Proof.} First we expose some auxiliary facts.

\begin{lemma}
\label{l3}
\begin{equation}
\label{OM2} \Omega=A^{-1} B = 1 - \frac{2}{\omega}\cdot
\partial_s^{-1} \cdot \omega_s.
\end{equation}
\end{lemma}

{\sc Proof of Lemma.} It is enough to show that $B = A\left(1 -
\frac{2}{\omega}\cdot \partial_s^{-1} \cdot \omega_s\right)$. Let us
write down the right-hand side of this equality as
$$
A - \omega^{-1}\cdot\partial_s\cdot\omega\frac{2}{\omega}\cdot
\partial_s^{-1} \cdot \omega_s=\omega^{-1}\cdot\partial_s\cdot\omega
- 2\omega^{-1}\omega_s=
$$
$$
= \omega^{-1}(\omega_s +\omega\cdot\partial_s) -
2\omega^{-1}\omega_s= -\omega^{-1}\omega_s + \partial_s=B.
$$
This proves the lemma.

\begin{corollary}
\label{sled4}
\begin{equation}
\label{OMx} (A^{-1} B)_r = \left(1 - \frac{2}{\omega}\cdot
\partial_s^{-1}\cdot \omega_s\right)_r=
\frac{2\omega_r}{\omega^2}\cdot \partial_s^{-1}\cdot  \omega_s -
\frac{2}{\omega}\cdot \partial_s^{-1}\cdot \omega_{rs}.
\end{equation}
\end{corollary}

Now we are ready to prove Theorem~\ref{th1}. It to enough to
establish the equivalent equality
\begin{equation}
\label{sopr2} \widetilde H \cdot \Omega =
\partial_s\cdot \Omega\cdot \partial_s^{-1}\cdot  H,
\end{equation}
i.e.
\begin{equation}
\label{sopr3} (\partial_r\partial_s - \widetilde u) \cdot A^{-1}
\cdot B = \partial_s \cdot A^{-1} \cdot B \cdot
\partial_s^{-1}\cdot  H.
\end{equation}
By Lemma \ref{l3}, the left-hand side of (\ref{sopr3}) is
$$
(\partial_r\partial_s - \widetilde u)\cdot A^{-1} \cdot B =
\partial_s\Big[
(A^{-1} B)_r + A^{-1} B\cdot \partial_r\Big]
-\left(-u+\frac{2\omega_r\omega_s}{\omega^2}\right)\cdot\left(1 -
\frac{2}{\omega}\cdot \partial_s^{-1}\cdot \omega_s\right)={}
$$
$$
{}= \Big[\partial_s \cdot\left(\frac{2\omega_r}{\omega^2}\cdot
\partial_s^{-1} \cdot  \omega_s
       - \frac{2}{\omega}\cdot \partial_s^{-1}\cdot \omega_{rs}\right) +
\partial_s\cdot A^{-1} B\cdot \partial_r\Big]+{}
$$
$$
{}+u\cdot \left(1 - \frac{2}{\omega}\cdot \partial_s^{-1}\cdot
\omega_s\right) -
 \frac{2\omega_r\omega_s}{\omega^2}\left(1 -
\frac{2}{\omega}\cdot \partial_s^{-1}\cdot \omega_s\right) ={}
$$
$$
{}= \left(\left(\frac{2\omega_r}{\omega^2}\right)_s \cdot
\partial_s^{-1}\cdot \omega_s
 + \frac{2\omega_r}{\omega^2}\cdot \partial_s
\cdot \partial_s^{-1}\cdot \omega_s \right.
\left. - \left(\frac{2}{\omega}\right)_s \cdot \partial_s^{-1}\cdot
\omega_{rs}
 - \frac{2}{\omega}\cdot
\partial_s\partial_s^{-1}\cdot \omega_{rs}\right) +{}
$$
$$
{}+
\partial_s\cdot A^{-1} B\cdot \partial_r
+u - \frac{2u}{\omega}\cdot \partial_s^{-1}\cdot \omega_s -
 \frac{2\omega_r\omega_s}{\omega^2} +
\frac{4\omega_r\omega_s}{\omega^3}\cdot \partial_s^{-1}\cdot
\omega_s ={}
$$
$$
{}= \left(\frac{2\omega_{rs}}{\omega^2}\cdot \partial_s^{-1}\cdot
\omega_s - \frac{4\omega_r\omega_s}{\omega^3}\cdot
\partial_s^{-1}\cdot \omega_s
 + \frac{2\omega_r \omega_s}{\omega^2} +
\left(\frac{2\omega_s}{\omega^2}\right)\cdot
\partial_s^{-1}\cdot \omega_{rs}  - \frac{2}{\omega} \omega_{rs}\right)+{}
$$
$$
{}+ \partial_s\cdot A^{-1} B\cdot \partial_r +u -
\frac{2u}{\omega}\cdot \partial_s^{-1}\cdot \omega_s -
 \frac{2\omega_r\omega_s}{\omega^2} + \frac{4\omega_r\omega_s}{\omega^3}\cdot
\partial_s^{-1}\cdot \omega_s
={}
$$
$$
{}=\left(\frac{2\omega_s}{\omega^2}\right)\cdot \partial_s^{-1}\cdot
(u\omega) - 2u +
\partial_s\cdot A^{-1} B\cdot \partial_r + u ={}
$$
$$
{}=\left(\frac{2\omega_s}{\omega^2}\right)\cdot \partial_s^{-1}\cdot
(u\omega) +
\partial_s\cdot A^{-1} B\cdot \partial_r - u .
$$
In the right-hand side of (\ref{sopr3}) we have
$$
\partial_s\cdot A^{-1} B\cdot\partial_s^{-1}\cdot(\partial_r\partial_s - u)=
 \partial_s\cdot A^{-1} B\cdot\partial_r -
 \partial_s\cdot\left(1 - \frac{2}{\omega}\cdot
\partial_s^{-1}\cdot \omega_s\right)\partial_s^{-1} \cdot u={}
$$
$$
{}=  \partial_s\cdot A^{-1} B\cdot\partial_r - u +
\left(\frac{2}{\omega}\right)_s\cdot \partial_s^{-1} \cdot
\omega_s\cdot \partial_s^{-1} \cdot u
 + \left(\frac{2}{\omega}\right)\cdot \partial_s\cdot
\partial_s^{-1}\cdot \omega_s\cdot \partial_s^{-1} \cdot u={}
$$
$$
{}=  \partial_s\cdot A^{-1} B\cdot\partial_r - u -
\frac{2\omega_s}{\omega^2}\cdot
\partial_s^{-1}\cdot \omega_s\cdot \partial_s^{-1} \cdot u
 + \frac{2\omega_s}{\omega}\cdot \partial_s^{-1} \cdot u.
$$
Let us apply Proposition~\ref{commutator} to the third term
$\partial_s^{-1}\cdot \omega_s\cdot \partial_s^{-1}  = \omega\cdot
\partial_s^{-1} - \partial_s^{-1} \cdot \omega$ in the last formula
and finally derive
$$
\partial_s\cdot A^{-1} B\cdot\partial_r - u -
 \frac{2\omega_s}{\omega^2}\cdot\left(\omega\cdot \partial_s^{-1} -
\partial_s^{-1} \cdot \omega\right) \cdot u
 + \frac{2\omega_s}{\omega}\cdot \partial_s^{-1} \cdot u={}
$$
$$
{}=\partial_s\cdot A^{-1} B\cdot\partial_r - u +
 \frac{2\omega_s}{\omega^2}\cdot \partial_s^{-1} \cdot \omega \cdot u,
$$
which coincides with the left-hand side of (\ref{sopr3}).

Theorem is proved.

Let us assume that $H$ is a (weakly) algebraic operator, i.e., there
are differential operators $L_1$ and $L_2$ such that
\begin{equation}
\label{Kri1} [L_1,L_2] = D_0 H, \ \ \ [L_1,H] = D_1 H, \ \ \ [L_2,H]
= D_2 H
\end{equation}
and there is a polynomial $Q$ in two variables with constant
coefficients such that
$$
Q(L_1,L_2) = 0\ (\mod H).
$$
By applying the Euclid algorithm (division with remainder in the
ring of formal linear \emph{ordinary} differential operators
$F(\partial_s)[\partial_r]$) we obtain $R_1, R_2 \in F(\partial_s)$
such that
\begin{equation}
\label{L2R} L_1 - Q_1\cdot M=R_1 \in F(\partial_s), \quad L_2 -
Q_2\cdot M=R_2 \in F(\partial_s)
\end{equation}
with $Q_i \in F(\partial_s)[\partial_r]$.

\begin{theorem}\label{th2}
\begin{enumerate}
\item
$[R_1,R_2] = 0$;

\item
$Q(R_1,R_2) = 0$;

\item
$[R_1,M] = [R_2,M]=0$.
\end{enumerate}
\end{theorem}

{\sc Proof.}
\begin{enumerate}
\item
Using (\ref{Kri1}) we compute that
$$
[R_1,R_2]=[L_1-Q_1 H,L_2-Q_2 H]=S_1\cdot H, \ \ \ S_1 \in
F(\partial_s)[\partial_r].
$$
However in the left-hand side we have an element of
$F(\partial_s)[\partial_r]$ which contains no derivations in $r$
which implies that $S_1=0$.

\item
The equality $Q(L_1,L_2) = 0\ (\mod H)$ means that $Q(L_1,L_2) = U
\cdot H$ in $F[\partial_r,\partial_s]$ which, by (\ref{Kri1}),
implies
$$
Q(R_1,R_2) = S_2 \cdot H \ \ \ \in F(\partial_s)[\partial_r].
$$
However in the left-hand side we have an element of
$F(\partial_s)[\partial_r]$ with no derivations in $r$ which implies
$S_2 = 0$.

\item
We have
$$
[R_1,H]=[L_1-Q_1 H, H]=S_3\cdot H,
$$
$$
[R_1,M]=[R_1,(\partial_s)^{-1}H]=[R_1,(\partial_s)^{-1}]H+
(\partial_s)^{-1}[R_1,H]=S_4\cdot H=S_5\cdot M,
$$
where $S_i \in F(\partial_s)[\partial_r]$ for $i=3,4,5$. By
Corollary~\ref{cor1}, this implies
$$
[R_1,M]=[R_1,\partial_r - (\partial_s)^{-1}\cdot u(r,s)]=
 -(R_1)_r + [R_1,(\partial_s)^{-1}\cdot u(r,s)] \in
F(\partial_s),
$$
and again as above we see that  $S_5=0$. The proof of the equality
$[R_2,M]=0$ is completely the same.
\end{enumerate}
Theorem is proved.

In view of Theorems~\ref{th1} and \ref{th2} we guess that the
Moutard transformation should preserve the class of weakly algebraic
two-dimensional Schr\"odinger operators.

\section{Applications of the Moutard transformation}

In \cite{TT07,TT08,TT-DAN08} we gave a simple examples of fast
decaying rational potentials for the two-dimensional Schr\"odinger
operator which has a degenerated $L_2$-kernel. These examples are
constructed by using the Moutard transformation as follows.

{\sc Main construction.} Let
$$
H_0 = - \Delta = -\Delta + u_0
$$
be an operator with a potential $u_0(x,y)$ and let $\omega_1(x,y)$
and $\omega_2(x,y)$ satisfy the equations
$$
H_0 \omega_1 = H_0 \omega_2 = 0.
$$
We take the Moutard transformations $M_{\omega_1}$ and
$M_{\omega_2}$ defined by $\omega_1$ and $\omega_2$ and obtain the
operators
$$
H_1 = -\Delta + u_1, \ \ \ H_2 = -\Delta + u_2
$$
where $u_1 = M_{\omega_1}(u_0), u_2 = M_{\omega_2}(u_0)$. By the
construction, we have
$$
H_1 M_{\omega_1}(\omega_2) = 0, \ \ \ H_2 M_{\omega_2}(\omega_1) =
0.
$$
Let us choose some function
$$
\theta_1 \in M_{\omega_1}(\omega_2)
$$
and put
$$
\theta_2  = -\frac{\omega_1}{\omega_2} \theta_1 \in
M_{\omega_2}(\omega_1).
$$
These functions define the Moutard transformations of $H_1$ and
$H_2$ and we obtain the operators $H_{12}$ and $H_{21}$ with the
potentials
$$
u_{12} = M_{\theta_1}(u_1), \ \ \ \ u_{21} = M_{\theta_2}(u_2).
$$
The following key lemma is checked by straightforward computations
which we omit.

\begin{lemma}
\begin{enumerate}
\item
$u_{12}=u_{21} = u$;
\item
For $\psi_1 = \frac{1}{\theta_1}$ and $\psi_2 = \frac{1}{\psi_2}$ we
have
$$
H\psi_1 = H\psi_2 = 0
$$
where $H = -\Delta+u$.
\end{enumerate}
\end{lemma}

We note that in this construction we have a free scalar parameter
$t$ (see (\ref{theta-family})) for the choice of $\theta_1 \in
M_{\omega_1}(\omega_2)$. This parameter can be used in some cases to
build a non-singular potential $u$ and functions $\psi_1$ and
$\psi_2$.

For example if we apply this construction to the situation when $u_0
= 0$ and $\omega_1$ and $\omega_2$ are real-valued harmonic
polynomials
$$
\omega_1 =  x +2(x^2-y^2)+xy, \ \ \ \omega_2 =
x+y+\frac{3}{2}(x^2-y^2)+5xy,
$$
then for some appropriate constant of integration in $\theta_1$ we
obtain
\begin{equation}
\label{ord2-1} u = -\frac{5120    (1 + 8    x + 2y + 17 x^2 + 17
y^2)}{(160 + 4    x^2 + 4y^2 + 16    x^3  + 4    x^2y + 16    x y^2
+ 4    y^3 + 17(x^2+y^2)^2)^2} =
\end{equation}
$$
= -\frac{5120|1+(4-i)z|^2}{(160+|z|^2|2+(4-i)z|^2)^2}
$$
and
\begin{equation}
\label{ord2-2}
\begin{split}
\psi_1 = \frac{x + 2    x^2 + x    y - 2    y^2}{160 + 4    x^2 +
4y^2 +  16    x^3  + 4    x^2    y + 16    x    y^2 + 4
y^3 + 17 (x^2+y^2)^2}, \\
\psi_2 = \frac{2    x + 2y + 3    x^2 + 10    x    y - 3    y^2}{160
+ 4    x^2 + 4y^2 +  16    x^3  + 4    x^2    y + 16    x    y^2 + 4
y^3 + 17 (x^2+y^2)^2}
\end{split}
\end{equation}
(here we simplify $\psi_1$ and $\psi_2$ by multiplying by some
constant).

\begin{theorem}[\cite{TT07,TT08}]
The potential $u$ given by (\ref{ord2-1}) is smooth, rational,  and
decays like $1/r^6$ for $r \to \infty$ (here $r = \sqrt{x^2+y^2}$).

The functions $\psi_1$ and $\psi_2$ given by (\ref{ord2-2}) are
smooth, rational, decay like $1/r^2$ for $r \to \infty$ and span a
two-dimensional space in the kernel of the operator $L = -\Delta +u:
L_2(\R^2) \to L_2(\R^2)$.
\end{theorem}

If one takes appropriate two harmonic polynomials of the third order
$\omega_1$ and $\omega_2$ (cf.\ \cite{TT07,TT08}) then one can
construct the potential $u$ and the functions $\psi_1$ and $\psi_2$
which are smooth, rational,  the potential decays like $1/r^8$ for
$r \to \infty$, the functions $\psi_1$ and $\psi_2$ decay like
$1/r^3$ for $r \to \infty$ and span a two-dimensional space in the
kernel of the operator $L = -\Delta +u: L_2(\R^2) \to L_2(\R^2)$.

{\sc Remark.} We guess that for every $N > 0$ by applying this
construction to other harmonic polynomials one can construct smooth
rational potentials $u$ and the eigenfunctions $\psi_1$ and $\psi_2$
decaying faster than $\frac{1}{r^N}$.

In \cite{TT08,TT-DAN08} we used a time-dependent extension of the
Moutard transformation constructing explicit solutions of the
Novikov--Veselov equation \cite{NV1,NV2}
\begin{equation}
\label{nv}
\begin{split}
U_t = \partial^3 U + \bar{\partial}^3 U + 3\partial(VU) +
3\bar{\partial} (\bar{V}U) =0,
\\
\bar{\partial}V = \partial U.
\end{split}
\end{equation}
Some of our solutions show a very special behavior: the initial data
for $t=0$ are smooth decaying rational functions of $x$ and $y$;
nevertheless for $t \geq t_0>0$ the solutions to the NV
equation~(\ref{nv}) blow up (become singular).

In particular the following solution $U(z,\bar{z},t)$ of the
Novikov--Veselov equation can be obtained using this technology:
$$
U= \frac{H_1}{H_2},
$$
with
$$
\begin{array}{rcl}
  H_1 &=&-12\Big(24 t x^2 + 12 t x + 24 t y^2 + 12 t y + x^5 - 3 x^4 y + 2
x^4 - 2 x^3 y^2 - 4 x^3 y \\
& & \ \ \ \ \ \ \ \ - 2 x^2 y^3 -  60 x^2 - 3 x y^4 - 4 x y^3 - 30 x
+ y^5 + 2 y^4 - 60 y^2 - 30 y\Big),
\end{array}
$$
$$ H_2=(3 x^4 + 4 x^3
+ 6 x^2 y^2 + 3 y^4 + 4 y^3 + 30 - 12 t)^2.
$$
This solution decays as $r^{-3}$ at infinity, it is nonsingular for
$0\leq t < T_\ast = \frac{29}{12}$ and have singularities for $t\geq
T_\ast = \frac{29}{12}$.

For $t \to T_\ast$, as we can see on Figure~1, the solution
$U(x,y,t)$ oscillates with growing amplitudes in the neighborhoods
of the points $P = (-1,0)$ and $Q = (0,-1)$ since the denominator
$H_2$ vanishes at these points for $t= T_\ast$. The numerator $H_1$
has zeros or order 3 at the points  $P$ and $Q$ so their respective
neighborhoods are subdivided by smooth lines into 6 sectors with
different signs of the numerator. The complicated behavior of the
potential in the neighborhood of one of the singular points for $t =
T_\ast$ is shown on Figure~2.

\unitlength 0.5mm \linethickness{0.5pt}
\begin{picture}(64.67,130.00)
\put(-20.67,17.00){\includegraphics[width=0.6\textwidth]{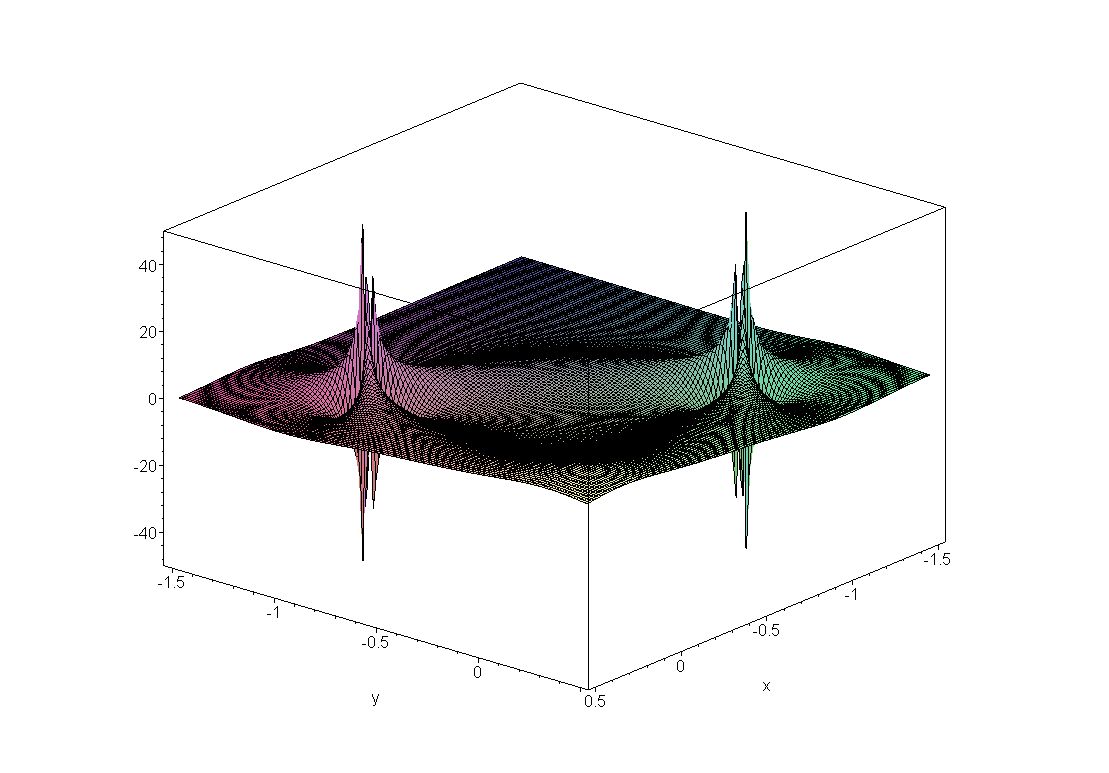}}
\put(5.67,5.00){\makebox(0,0)[lc]{Figure 1: The potential $U$ for $t
= \frac{29}{12}$.}}
\put(120.67,17.00){\includegraphics[width=0.6\textwidth]{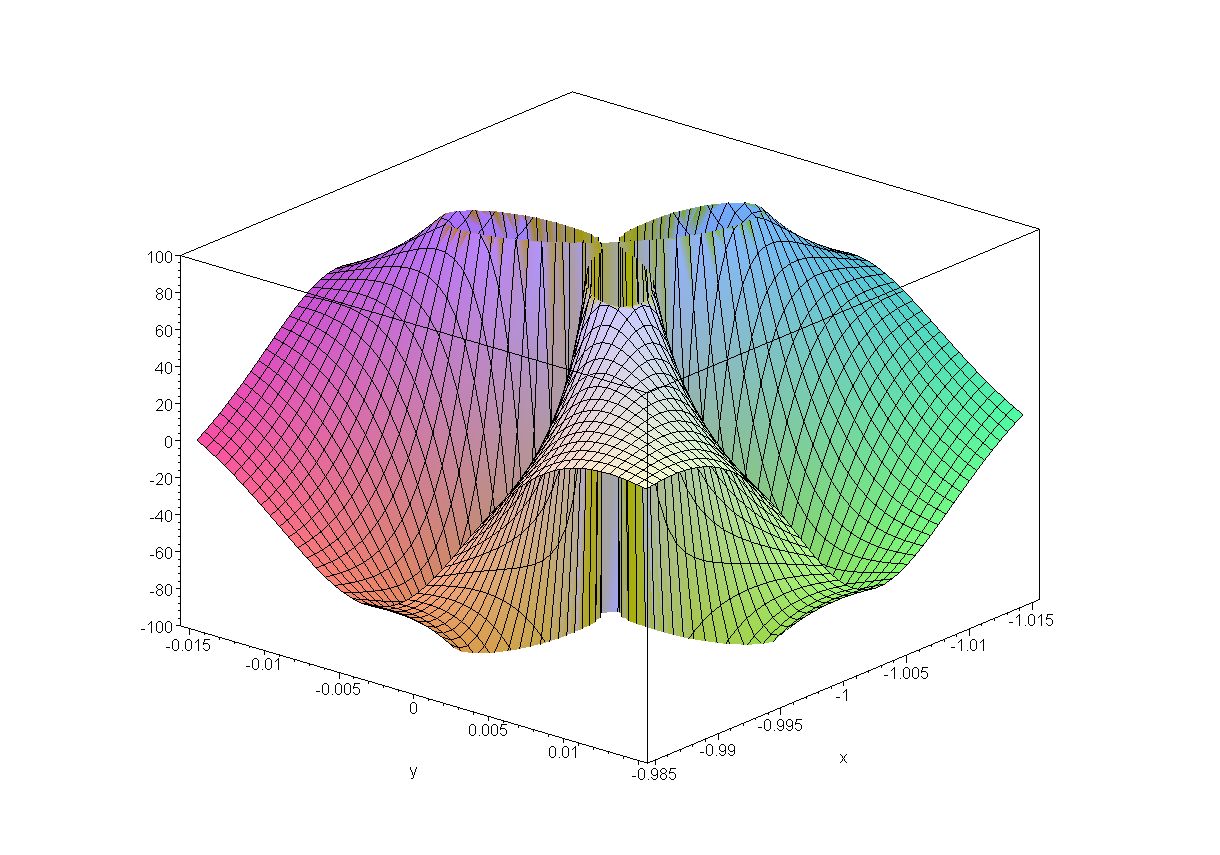}}
\put(145.67,5.00){\makebox(0,0)[lc]{Figure 2: The potential $U$ for
$t = \frac{29}{12}$}}
 \put(133.67,-5.00){\makebox(0,0)[lc]{ in the neighborhood of the point $(-1,0)$.}}
\end{picture}


\begin{thebibliography}{MMM}


\bibitem{AMM}
\textsc{H. Airault, H.P. McKean and J. Moser}, Rational and elliptic
solutions of the Korteweg-de Vries equation and a related many-body
problem, Comm. Pure Appl. Math. {\bf 30} (1977), 95--148.

\bibitem{BC}
\textsc{J.L. Burchnall and T.W. Chaundy}, Commutative ordinary
differential operators, Proc. London Soc., Ser. 2, {\bf 21} (1923),
420--440.

\bibitem{Crum}
\textsc{M.M. Crum}, Associated Sturm--Liou\-vil\-le sys\-tems,
Quart. J. Math., Ser. 2 {\bf 6} (1955), 121--127.

\bibitem{Dar}
\textsc{G. Darboux}, Le\c{c}ons sur la th\'eorie g\'en\'erale des
surfaces et les applications g\'eom\'etriques du calcul
infinit\'esimal. T.~2. Paris: Gautier-Villars, 1915.


\bibitem{DKN}
\textsc{B.A. Dubrovin, I.M. Krichever and S.P. Novikov}, The
Schr\"o\-din\-ger equation in a periodic field and Riemann surfaces,
Soviet Math. Dokl. {\bf 17} (1976), 947--952.

\bibitem{DMN}
\textsc{B.A. Dubrovin, V.B. Matveev and S.P. Novikov}, Nonlinear
equations of Korteweg--de Vries type, finite-zone linear operators,
and Abelian varieties, Russian Math. Surveys {\bf 31}:1 (1976),
59--146.

\bibitem{EK}
\textsc{F. Ehlers and H. Kn\"orrer}, An algebro-geometric
interpretation of the B\"acklund transformation for the Korteweg--de
Vries equation, Comm. Math. Helv. {\bf 57} (1982), 1--10.

\bibitem{K76}
\textsc{I.M. Krichever}, Methods of algebraic geometry in the theory
of non-linear equations, Russian Math. Surveys {\bf 32}:6 (1977),
185--213.

\bibitem{mou}
\textsc{Th. Moutard},
 Sur la construction des \'equations de la forme
$\frac{1}{z} \frac{d^2 z}{d x\, d y} = \lambda(x,y)$, qui admettent
une int\'egrale g\'en\'erale explicite, J. \'Ecole Polytechnique.
{\bf 45} (1878), 1--11.

\bibitem{N}
\textsc{S.P. Novikov}, A periodic problem for the Korteweg-de Vries
equation. I, Functional Anal. Appl. {\bf 8} (1974), 236--246.

\bibitem{NV1}
\textsc{S.P. Novikov and A.P. Veselov}, Finite-zone, two-dimensional
Schr\"odinger operators. Potential operators, Soviet Math. Dokl.
{\bf 30} (1984), 705--708.

\bibitem{NV2}
\textsc{S.P. Novikov and A.P. Veselov}, Finite-zone, two-dimensional
potential Schr\"odinger operators. Explicit formulas and evolution
equations, Soviet Math. Dokl. {\bf 30} (1984), 588--591.

\bibitem{NV}
\textsc{S.P. Novikov and A.P. Veselov}, Exactly solvable
two-dimensional Schr\"o\-dinger operators and Laplace
transformations, Translations of the Amer. Math. Soc., Ser. 2, {\bf
179} (1997), 109--132.

\bibitem{Ore}
\textsc{O. Ore}, Linear equations in non-commutative fields, Ann. of
Math. {\bf 32} (1931), 463--477.

\bibitem{TT07}
\textsc{I.A. Taimanov and S.P. Tsarev}, Two-dimen\-sio\-nal
Schr\"odinger operators with fast decaying rational potential and
multidimensional $L_2$-kernel, Russian Mathematical Surveys, {\bf
62}:3 (2007), 631-633.

\bibitem{TT08}
\textsc{I.A. Taimanov and S.P. Tsarev}, Two-dimen\-sio\-nal rational
solitons and their blowup via the Moutard transformation,
Theoretical and Mathematical Physics, {\bf 157}:2 (2008),
1525--1541.

\bibitem{TT-DAN08}
\textsc{I.A. Taimanov and S.P. Tsarev}, Blowing up solutions of the
Novikov--Veselov equation, Doklady Mathematics,  {\bf 77}:3 (2008),
467--468.



\end{thebibliography}
\end{document}